\documentclass[acmsmall]{acmart}

\AtBeginDocument{%
  \providecommand\BibTeX{{%
    \normalfont B\kern-0.5em{\scshape i\kern-0.25em b}\kern-0.8em\TeX}}}

\setcopyright{acmcopyright}
\copyrightyear{2022}
\acmYear{2022}
\acmDOI{XXXXXXX.XXXXXXX}

\acmConference[PEARC '22]{Make sure to enter the correct
  conference title from your rights confirmation email}{July 10--14,
  2022}{Boston, MA}
\begin{document}

\title[Cyberinfrastructure value]{Cyberinfrastructure value: a survey on perceived importance and usage}

\author{Praneeth Chityala}
\email{pkchitya@iu.edu}
\orcid{0000-0001-6139-622X}
\affiliation{%
  \institution{Pervasive Technology Institute, Indiana University}
  \streetaddress{2709 E. 10th St.}
  \city{Bloomington}
  \state{Indiana}
  \country{USA}
  \postcode{47408}
}

\author{Claudia M. Costa}
\affiliation{
  \institution{Pervasive Technology Institute, Indiana University}
  \city{Bloomington}
  \state{Indiana}
  \country{USA}}
\email{clcosta@iu.edu}
\orcid{0000-0002-9869-1780}

\author{Julie A. Wernert}
\affiliation{
  \institution{Pervasive Technology Institute, Indiana University}
  \city{Bloomington}
  \state{Indiana}
  \country{USA}}
\email{jwernert@iu.edu}
\orcid{0000-0002-5705-9527}

\author{Craig A. Stewart}
\orcid{0000-0003-2423-9019}
\affiliation{%
  \institution{National Center for Supercomputing Applications, University of Illinois at Urbana-Champaign}
  \city{Urbana}
  \state{IL}
  \country{USA}
}
\affiliation{%
  \institution{Department of Computer Science, Indiana University}
  \city{Bloomington}
  \state{IN}
  \country{USA}
}

\renewcommand{\shortauthors}{Chityala et al.}

\begin{abstract}
The research landscape in science and engineering is heavily reliant on computation and data storage. The intensity of computation required for many research projects illustrates the importance of the availability of high performance computing (HPC) resources and services. This paper summarizes the results of a recent study among principal investigators that attempts to measure the impact of the cyberinfrastructure (CI) resources allocated by the XSEDE (eXtreme Science and Engineering Discovery Environment) project to various research activities across the United States. Critical findings from this paper include: a majority of respondents report that the XSEDE environment is important or very important in completing their funded work, and two-thirds of our study’s respondents developed products (e.g., datasets, websites, software, etc.) using XSEDE-allocated resources. With nearly one-third of respondents citing the importance of XSEDE-allocated resources in securing research funding, we estimate that respondents of this survey have secured approximately \$3.3B in research funding from various sources, as self-reported by respondents.

\end{abstract}






\maketitle

\section{Introduction}
Across the world, increasingly tight government and university budgets prompt increasingly stringent requirements for scientific research funding. Similar concerns persist in private-sector research and development organizations, which demand early evidence of viability and profitability. This issue is particularly acute for higher education institutions in the United States due to ever-increasing competition for federal research funding, coupled with financial pressures related to projected decreases in university enrollments \cite{stewart2019financial}. Helping to bridge this gap, XSEDE is an organization that facilitates, allocates, and supports access to NSF-funded and other cyberinfrastructure, providing researchers secure computational and data storage resources required for their research. Such resources and services (e,g,, supercomputers, collections of data, and computational tools) aid innovation and discovery \cite{XSEDEweb}.

This paper uses survey methods to ascertain the value of XSEDE to those who use its services, focusing on the opinions of Principal Investigators (PIs) who have received a resource allocation from September 1, 2016, to August 31, 2021. The majority of these allocations are for the use of computing systems funded by the NSF through separate funding actions (i.e., independent of XSEDE) to Service Providers. Some PIs also received allocations of consulting and programming effort directly from XSEDE. Thus, the services investigated are a mix of system, software, and application services, with a large portion of the study related to running software applications on XSEDE-allocated resources. To better understand the relationship between success in competing for grant awards and plans for executing the activities described in the funded proposals, the study inquired about the use of advanced CI resources allocated by, provided by, or accessed via the XSEDE CI environment. Therefore the content of this paper, while of general interest to PEARC attendees, seems most closely related to the Software and Applications Track.

\section{Methods}
This study was reviewed by the Indiana University (IU) Institutional Review Board prior to its initiation and was determined to be “expedited,” meaning there was nothing sensitive in the questions asked. All participants gave informed consent prior to taking part in the study. The study population was identified from XSEDE allocation data and consisted of PIs who received resource allocations during XSEDE2 program years, beginning 2016 through 2021. Only PIs with projects classified as "Research" or "XSEDE Rapid Response" (a special allocation category for COVID research) were included. The survey was conducted via email with personalized links. Population members received an initial letter of invitation on January 13, 2022, followed by four reminder messages, the last of which being on February 4, 2022. The questionnaire consisted of 7-20 questions, depending on answers to screening questions. After accounting for bounced or undeliverable emails, the effective survey population consisted of 1595 members. Of these, 700 agreed to participate and went on to complete or partially answer questions in the survey. The effective response rate was ~44\% (700 out of 1595)\cite{morton2012survey}. Such a high response rate speaks to the importance PIs place on the value of XSEDE in their research activities.

\section{Results}
\subsection{Importance of XSEDE in securing funds and completing funded work}

\begin{figure}[ht]
  \centering
  \includegraphics[width=\linewidth]{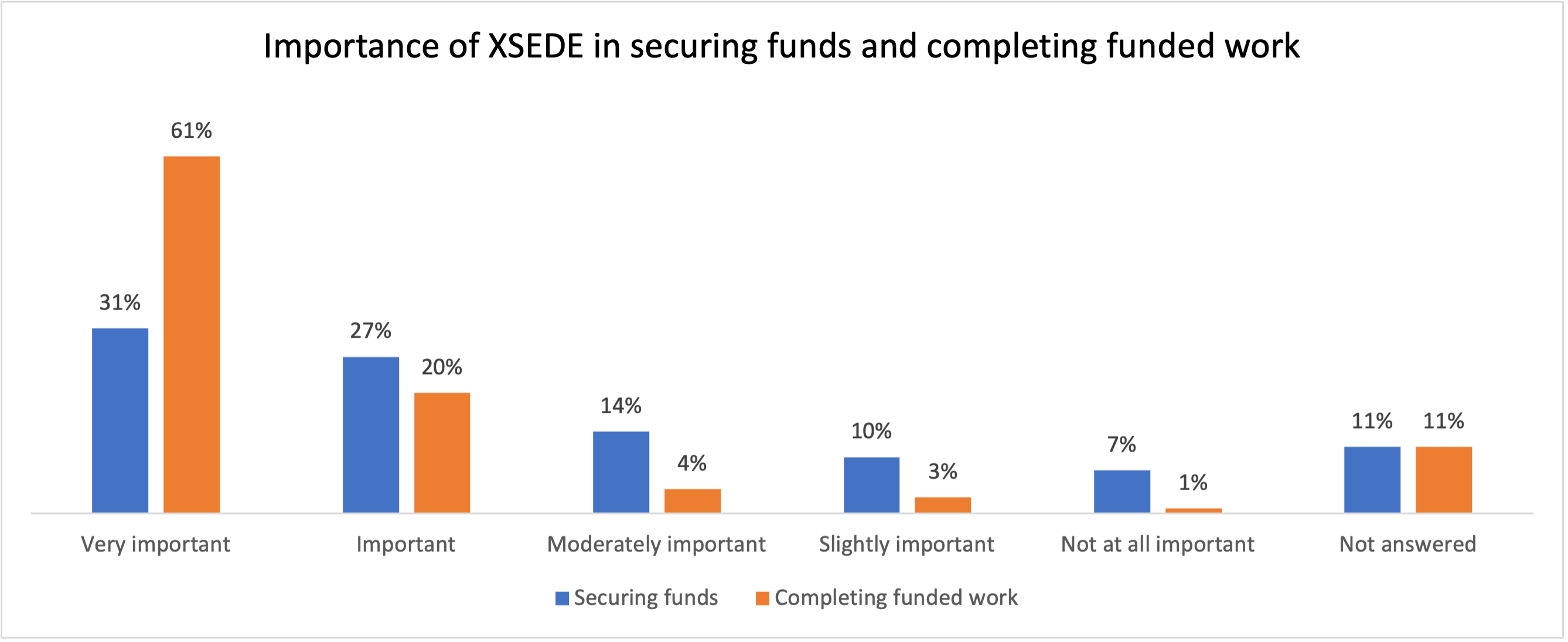}
  \caption{Importance of XSEDE in securing funds and completing funded work}
  \label{fig:FundedWork}
  \Description{-}
\end{figure}

The importance of XSEDE is tracked using two questions: What is the level of importance of XSEDE in 1) securing research funding and 2) completing funded work. The majority of respondents reported that XSEDE is important or very important in securing funds and completing funded work. Details are summarized in Fig. \ref{fig:FundedWork}.

Further analysis shows XSEDE resources are \emph{very important} for 61\% (424 of 700) of respondents in completing funded work, and for 31\% (220 of 700) for securing funds. One (1) percent (6 of 700) of respondents reported that XSEDE resources are \emph{not at all important} in completing funded work, and seven (7) percent (51 of 700) reporting XSEDE resources are \emph{not at all important} for securing funds. 

Another key finding is that 63\% (444 of 700) of respondents stated that the funded work could not have been completed without the availability of XSEDE-allocated resources. Only seven percent (49 of 700) of respondents said that they could have completed the funded work without XSEDE-allocated resources. 

In addition, XSEDE’s Extended Collaborative Support Services (ECSS), often described as expert consulting services and a key service offering, was cited by 61\% (106 of 175) of respondents who had used the service as either important or very important in completing their funded work.

\subsection{Usage analysis of XSEDE-allocated resources}

\begin{figure}[ht]
  \centering
  \includegraphics[width=\linewidth]{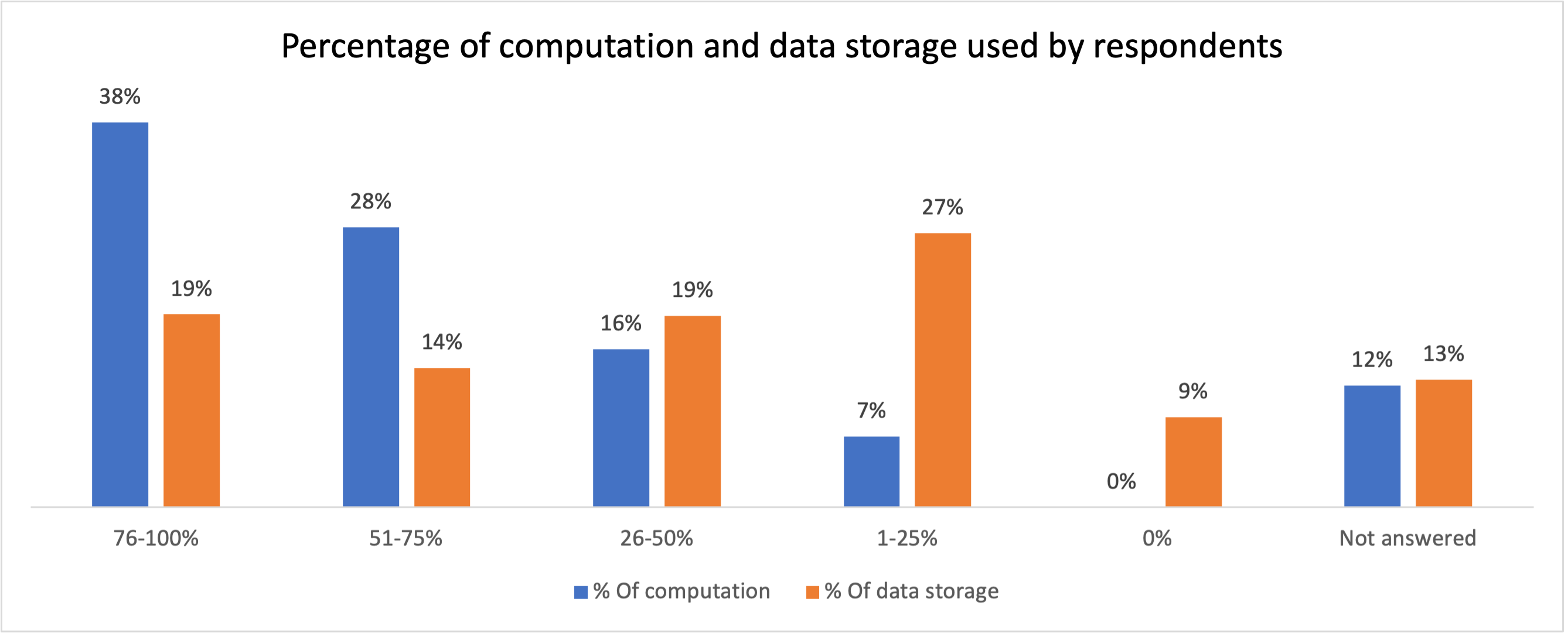}
  \caption{Percentage of computation and data storage provided by XSEDE}
  \label{fig:ResourcesUsed}
  \Description{-}
\end{figure}

XSEDE provides CI resources to support research activities in the two key areas of computational capabilities and  data storage. Our survey captured two important metrics for the projects using XSEDE-allocated resources: 1) the percentage of computation performed on computational resources, and 2) the percentage of data stored on data storage resources. These questions are aimed at understanding the role of XSEDE-allocated resources and the individual’s research needs. Details are summarized in Fig. \ref{fig:ResourcesUsed}.

Further analysis indicated that 38\% (265 of 700) of respondents used XSEDE-allocated resources for more than 75\% of their project’s computational needs, and 19\% (133 of 700) used XSEDE-allocated resources for more than 75\% of their project’s data storage requirements. No respondents reported having not used XSEDE-allocated resources for their project's computational needs; nine percent (62 of 700) of respondents indicated they had not used any XSEDE-allocated resources for their project's data storage requirements. It was also reported that more than 89\% (625 of 700) of respondents used their XSEDE allocation resources before they expired.

\subsection{Research products developed using XSEDE-allocated resources}

\begin{figure}[ht]
  \centering
  \includegraphics[width=\linewidth]{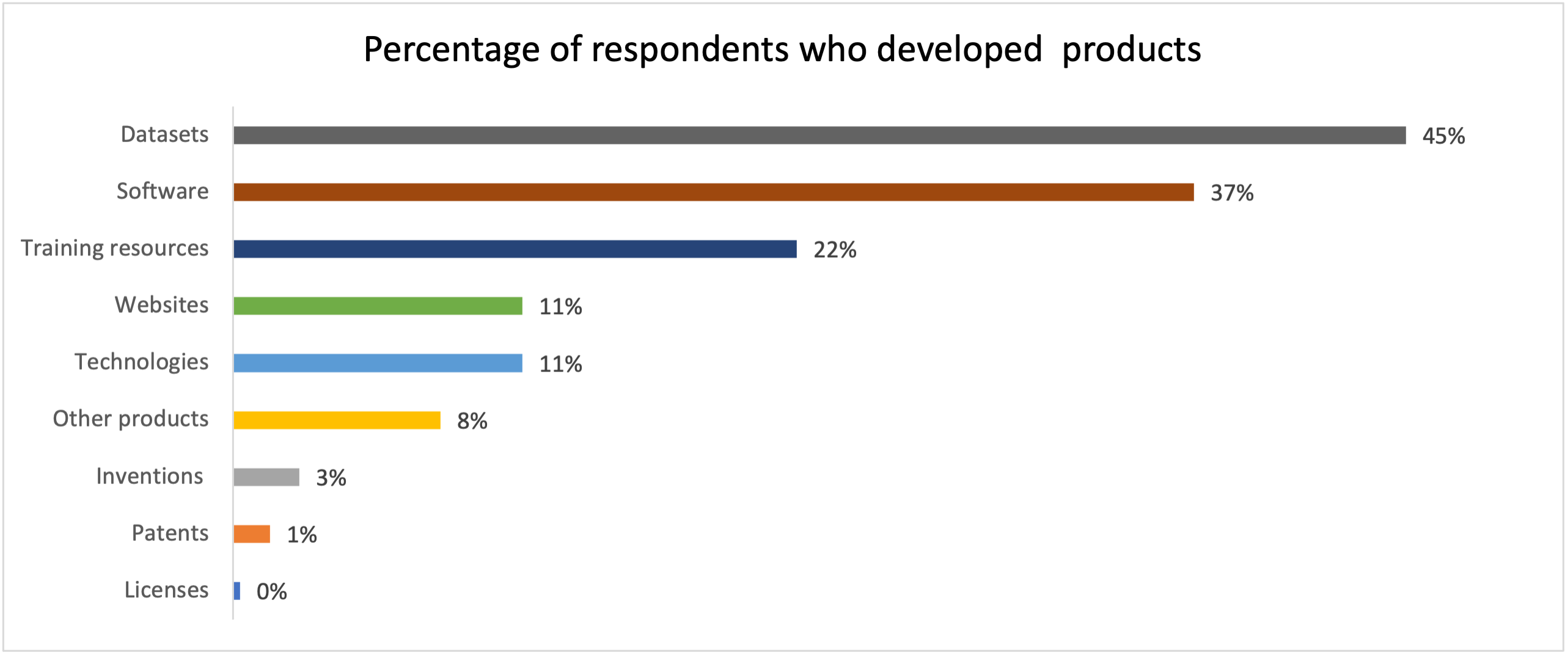}
  \caption{Percentage of respondents reporting that XSEDE was used to develop indicated products}
  \label{fig:ProductsReported}
  \Description{-}
\end{figure}

Our study also collected other important non-financial metrics, including the kinds of research products developed using XSEDE-allocated resources. More than 66\% (463 of 700) of responding PIs reported having developed products (in addition to publications) using XSEDE-allocated resources. Figure \ref{fig:ProductsReported} illustrates the research products developed using XSEDE-allocated resources.

Many responding PIs reported having developed more than one kind of product, with 45\% (316 of 700) having developed datasets and 37\% (259 of 700) having developed software. These results exemplify the non-financial returns resulting from the availability of XSEDE-allocated resources.

\subsection{Analysis across fields of science}

 The XSEDE allocations data provided individual research activities and domains for 700 respondents representing 737 projects (multiple PIs have allocations on more than one project). This data also had grant values that are self-reported by the researchers without the corresponding funding source. Corresponding grant values are available for only 495 of 700 responding PIs; grant values for the remaining 205 respondents are treated as zero dollars. 
 Based on that data, we looked at the distribution of resources and products developed per domain area. Engineering and Technology, Biological Sciences, and Physical Sciences command the greatest percentages of XSEDE resources in terms of computation and data storage and, in turn, report the greatest number of non-publication products produced. Further investigation shows that there is no significant difference between the distribution of products developed and resources used when compared with the distribution of respondents across domains. Table \ref{table:NonFinancialReturns} illustrates the CI usage and products developed over various domains.

\begin{table}[htbp]
\centering
\footnotesize
\caption{Distribution of CI usage and product development across different fields of science}
\label{table:NonFinancialReturns}
\begin{tabular}{ |p{4cm}|p{1.7cm}|p{1.3cm}|p{1.8cm}|p{1.1cm}|}
 \hline
\bf Field of Science & \hfil \bf Respondents &\hfil \bf Products &\hfil \bf Computation &\hfil \bf Storage \\
 \hline
Engineering and Technology   &   \hfil25\%   &   \hfil27\%   &  \hfil 26\%   &   \hfil27\% \\ 
Biological Sciences   &   \hfil23\%   &   \hfil21\%   &  \hfil 22\%   &   \hfil21\% \\ 
Physical Sciences   &   \hfil20\%   &   \hfil20\%   &  \hfil 19\%   &   \hfil21\% \\ 
Chemical Sciences   &   \hfil12\%   &   \hfil9\%   &  \hfil 12\%   &   \hfil11\% \\ 
Earth and Environmental Sciences   &   \hfil10\%   &   \hfil11\%   &  \hfil 10\%   &   \hfil10\% \\ 
Computer and Information Sciences   &   \hfil5\%   &   \hfil6\%   &  \hfil 5\%   &   \hfil5\% \\ 
Mathematics   &   \hfil2\%   &   \hfil2\%   &  \hfil 2\%   &   \hfil2\% \\ 
Social Sciences   &   \hfil2\%   &   \hfil1\%   &  \hfil 1\%   &   \hfil1\% \\ 
Other Natural Sciences   &   \hfil1\%   &   \hfil1\%   &  \hfil 1\%   &   \hfil1\% \\ 
Medical and Health Sciences   &   \hfil1\%   &   \hfil0\%   &  \hfil0\%   &   \hfil1\% \\ 
Other   &   \hfil1\%   &   \hfil0\%   &  \hfil1\%   &   \hfil1\% \\ 
 \hline
 
\end{tabular}
\end{table}


The total funding reported by the PIs was in excess of \$3.3B. About \$2.3B of that is related to the 63\% of PIs who said they wouldn't have been able to complete their funded work without XSEDE-allocated resources. After linear extrapolation to the total population of 1595 XSEDE-allocated PIs, the total funding stands in excess of \$7.5B. This speaks to the impact - financial, scientific, and otherwise - that the XSEDE project has had on the US scientific community. These grants are further distributed across fields of science as stated in Table \ref{table:GrantsoveFoS}.

\begin{table}[htbp]
\centering
\footnotesize
\caption{Grant distribution over fields of science}
\label{table:GrantsoveFoS}
\begin{tabular}{ |p{4cm}|p{2.3cm}| p{1.2cm}|}
 \hline
\bf Field of Science   &   \hfil \bf Grants   &   \hfil \bf \% Grants  \\ 
 \hline
Engineering and Technology   &   \hfil\$1,081,804,549   &   \hfil32.6\% \\ 
Biological Sciences   &   \hfil\$676,013,433   &   \hfil20.3\% \\ 
Physical Sciences   &   \hfil\$475,286,361   &   \hfil14.3\% \\ 
Chemical Sciences   &   \hfil\$425,359,814   &   \hfil12.8\%  \\ 
Earth and Environmental Sciences   &   \hfil\$390,404,665   &   \hfil11.7\%  \\ 
Computer and Information Sciences   &   \hfil\$183,310,173   &   \hfil5.5\%  \\ 
Medical and Health Sciences   &   \hfil\$61,388,425   &   \hfil1.8\%  \\
Other Natural Sciences   &   \hfil\$10,165,754   &   \hfil0.3\%  \\ 
Other   &   \hfil\$6,584,192   &   \hfil0.2\%  \\ 
Mathematics   &   \hfil\$7,642,390   &   \hfil0.2\%  \\ 
Social Sciences   &   \hfil\$4,825,860   &   \hfil0.1\%  \\
 \hline
 Total   &   \hfil\$3,322,785,616   &   \hfil100.0\%  \\ 
 \hline
 
\end{tabular}
\end{table}

The Engineering and Technology category accounts for a major share of the grants secured at 33\% of the total, whereas grants from Biological Sciences, Physical Sciences, Chemical Sciences, and Earth and Environmental Sciences range from 10\% to 20\%.

\section{Discussion}
The high response rate of 44\% suggests that PIs recognize the importance of the XSEDE resources they have used. Such a rate of response for a survey that did not provide incentives for completion is considered above average\cite{deutskens2004response}.

Survey outcomes suggest that XSEDE is providing a great deal of value to the work of researchers. Over 80\% of respondents said that XSEDE was \emph{important} or \emph{very important} in completing funded research. This speaks to the value XSEDE provides in supporting research that is regarded as "worthy" in peer-reviewed funding decisions made by federal agencies in particular.

Survey results are in some ways consistent with the general history of NSF-provided national CI resources, and in some ways different. Consistent with the origins of the NSF Supercomputing Centers program of the 1980s, survey results suggest that computation is still a more important service than storage. There are interesting differences between the usage of XSEDE-allocated resources found in this survey as compared to earlier analyses of usage. While Physics and Chemistry are traditionally considered  disciplines that use a great deal of NSF-provided resources, the highest numbers of responses came from those in Engineering and Technology and Biological Sciences, with these two disciplines accounting for almost half of our respondents. Reports of usage for the TeraGrid, the predecessor to XSEDE, are primarily in "amount of resource used," \cite{catlett2008teragrid} suggesting that the relative use of XSEDE by engineering and biology researchers is higher than for the TeraGrid.

The most frequent types of products reported were datasets and software. Interestingly, almost a third of the respondents indicated that they had developed some sort of training resource based on (or about) the use of XSEDE. Inventions were reported as one of the outcomes for three percent of the respondents. Patents represent a particularly valuable and potentially marketable product; one percent of respondents indicated that a patent had resulted from their use of XSEDE.

The value of funded research supported by XSEDE is interesting and important. Over 11 years of XSEDE the total budget is less than \$300M. The total amount of grant funding that survey respondents indicated had been supported by XSEDE was more than \$3B. 

\section{Conclusion}
XSEDE plays a crucial role in allocating required CI resources for many research activities across the United States. This paper has discussed the survey on usage of XSEDE-allocated CI resources over the 2016-2021 period, along with datasets that provide the (self-reported) grants secured by individual research activities and the division of research as per NSF. The high response rate of 44\% speaks to the enthusiasm of PIs for describing the benefits they received from XSEDE allocations. A large majority of respondents said XSEDE-allocated resources were important or very important in completing their funded work. About 63\% reported that they would not be able to complete their funded research without XSEDE-allocated resources. Nearly 66\% reported that they had developed products (e.g., websites, software, datasets, etc.) using these resources, testifying to the tangible outcomes of these resources. This implies the pivotal role of XSEDE-allocated CI resources in their research activities.

\begin{acks}
This work was supported in part by NSF award 1548562 (John Towns, PI) and by the IU Pervasive Technology Institute. Thanks to Ken Hackworth and Dave Hart for XSEDE allocations data. Thanks to Marlon Pierce for valuable comments and thanks to Kristol Hancock and Tonya Miles for editing. Any opinions presented here are those of the authors and may not reflect those of supporting organizations.
\end{acks}

\bibliographystyle{ACM-Reference-Format}
\bibliography{Bibliography}


\begin{thebibliography}{5}


\ifx \showCODEN    \undefined \def \showCODEN     #1{\unskip}     \fi
\ifx \showDOI      \undefined \def \showDOI       #1{#1}\fi
\ifx \showISBNx    \undefined \def \showISBNx     #1{\unskip}     \fi
\ifx \showISBNxiii \undefined \def \showISBNxiii  #1{\unskip}     \fi
\ifx \showISSN     \undefined \def \showISSN      #1{\unskip}     \fi
\ifx \showLCCN     \undefined \def \showLCCN      #1{\unskip}     \fi
\ifx \shownote     \undefined \def \shownote      #1{#1}          \fi
\ifx \showarticletitle \undefined \def \showarticletitle #1{#1}   \fi
\ifx \showURL      \undefined \def \showURL       {\relax}        \fi
\providecommand\bibfield[2]{#2}
\providecommand\bibinfo[2]{#2}
\providecommand\natexlab[1]{#1}
\providecommand\showeprint[2][]{arXiv:#2}

\bibitem[Deutskens et~al\mbox{.}(2004)]%
        {deutskens2004response}
\bibfield{author}{\bibinfo{person}{Elisabeth Deutskens}, \bibinfo{person}{Ko
  De~Ruyter}, \bibinfo{person}{Martin Wetzels}, {and} \bibinfo{person}{Paul
  Oosterveld}.} \bibinfo{year}{2004}\natexlab{}.
\newblock \showarticletitle{Response rate and response quality of
  internet-based surveys: an experimental study}.
\newblock \bibinfo{journal}{\emph{Marketing letters}} \bibinfo{volume}{15},
  \bibinfo{number}{1} (\bibinfo{year}{2004}), \bibinfo{pages}{21--36}.
\newblock


\bibitem[et. al(2008)]%
        {catlett2008teragrid}
\bibfield{author}{\bibinfo{person}{Catlett et. al}.}
  \bibinfo{year}{2008}\natexlab{}.
\newblock \showarticletitle{Teragrid: Analysis of organization, system
  architecture, and middleware enabling new types of applications}.
\newblock \bibinfo{journal}{\emph{High Performance Computing and Grids in
  Action}}  \bibinfo{volume}{16} (\bibinfo{year}{2008}),
  \bibinfo{pages}{225--249}.
\newblock
\urldef\tempurl%
\url{https://hdl.handle.net/2022/14524}
\showURL{%
\tempurl}


\bibitem[Morton et~al\mbox{.}(2012)]%
        {morton2012survey}
\bibfield{author}{\bibinfo{person}{Susan Morton}, \bibinfo{person}{Dinusha
  Bandara}, \bibinfo{person}{Elizabeth Robinson}, {and} \bibinfo{person}{Polly
  Atatoa-Carr}.} \bibinfo{year}{2012}\natexlab{}.
\newblock \showarticletitle{In the 21st Century, what is an acceptable response
  rate?}
\newblock \bibinfo{journal}{\emph{Australian and New Zealand journal of public
  health}}  \bibinfo{volume}{36} (\bibinfo{date}{04} \bibinfo{year}{2012}),
  \bibinfo{pages}{106--8}.
\newblock
\urldef\tempurl%
\url{https://doi.org/10.1111/j.1753-6405.2012.00854.x}
\showDOI{\tempurl}


\bibitem[Stewart et~al\mbox{.}(2019)]%
        {stewart2019financial}
\bibfield{author}{\bibinfo{person}{Craig~A Stewart}, \bibinfo{person}{David~Y
  Hancock}, \bibinfo{person}{Julie Wernert}, \bibinfo{person}{Thomas Furlani},
  \bibinfo{person}{David Lifka}, \bibinfo{person}{Alan Sill},
  \bibinfo{person}{Nicholas Berente}, \bibinfo{person}{Donald~F McMullen},
  \bibinfo{person}{Thomas Cheatham}, \bibinfo{person}{Amy Apon},
  {et~al\mbox{.}}} \bibinfo{year}{2019}\natexlab{}.
\newblock \showarticletitle{Assessment of financial returns on investments in
  cyberinfrastructure facilities: A survey of current methods}.
\newblock In \bibinfo{booktitle}{\emph{Proceedings of the Practice and
  Experience in Advanced Research Computing on Rise of the Machines
  (learning)}}. \bibinfo{publisher}{ACM}, \bibinfo{address}{Chicago, IL, USA},
  \bibinfo{pages}{1--8}.
\newblock
\urldef\tempurl%
\url{https://doi.org/10.1145/3332186.3332228}
\showURL{%
\tempurl}


\bibitem[XSEDE(2022)]%
        {XSEDEweb}
\bibfield{author}{\bibinfo{person}{XSEDE}.} \bibinfo{year}{2022}\natexlab{}.
\newblock \bibinfo{booktitle}{\emph{About XSEDE}}.
\newblock XSEDE.
\newblock
\urldef\tempurl%
\url{https://www.xsede.org/about/what-we-do}
\showURL{%
Retrieved April 1, 2022 from \tempurl}


\end{thebibliography}
 

\end{document}